\magnification=\magstep1
\centerline{\bf The Casimir Effect and Thermodynamic Instability}
\bigskip
\centerline{A. Widom [1], E. Sassaroli [1,2] Y.N. Srivastava [1-3], 
J. Swain [1]}
\centerline{1.Physics Department, Northeastern University, 
Boston MA 01225, USA}
\centerline{2. Laboratory for Nuclear Science and Department of Physics} 
\centerline {Massachusetts Institute of Technology, Cambridge, MA 02139, USA}
\centerline{3. Dipartimento di Fisica and Sezione INFN, Universita' 
di Perugia, I-06123 Perugia, Italy}
\bigskip
\bigskip
\centerline{\it ABSTRACT}
\bigskip
One loop field theory calculations of free energies quite often yield 
violations of the stability conditions usually associated 
with the thermodynamic second law. Perhaps the best known example 
involves the equation of state of black holes. Here, it is pointed out 
that the Casimir force between two parallel conducting plates also 
violates a thermodynamic stability condition normally associated with 
the second law of thermodynamics.  
\vskip .6in
\centerline{\bf 1. Introduction}
\bigskip

A property shared by many one loop quantum statistical thermodynamic 
computations is that a thermodynamic second law instability appears in 
the final answer. Perhaps the most commonly discussed example of this 
phenomena occurs in black hole statistical thermodynamics [1,2]. The entropy 
$S$ of a black hole having mass $M$ is given by 
$$
S=4\pi k_B\Big({GM^2\over \hbar c}\Big), \eqno(1.1)
$$
where $G$ is Newton's gravitational coupling strength. The black hole 
temperature, defined as $T=c^2(\partial M/\partial S)$, is then 
determined by 
$$
M=\Big({\hbar c^3 \over 8\pi G k_BT}\Big). \eqno(1.2)
$$ 
The black hole heat capacity $C=c^2(\partial M/\partial T)$ is 
thereby negative, 
$$
{C\over k_B}=-\Big({\hbar c^5 \over 8\pi G (k_BT)^2}\Big)
=-\Big({1\over 8\pi }\Big)
\Big({\hbar c\over k_BT \Lambda }\Big)^2<0, \eqno(1.3)
$$
which violates the second law of thermodynamics. In Eq.(1.3), 
$\Lambda $ is the Planck length. 

That a one loop quantum gravity calculation, i.e. the gravitational 
Casimir effect, produces a theoretical violation of the 
thermodynamic second law is (perhaps) not very surprising. 
Even at the Newtonian theoretical level, the long 
range gravitational attraction upsets the usual second law 
convexity conditions otherwise present in the thermodynamic 
limit of infinite size. 

Our purpose is to show the perhaps more surprising result that 
the electrodynamic Casimir effect can also produce a second law 
violation, which should be present in laboratory experiments. To 
see what is involved, suppose that a material is located inside 
a box of volume $V$. The free energy obeys the thermodynamic 
law 
$$
dF=-SdT-PdV. \eqno(1.4)
$$
If the box is a cylinder cavity with a cross sectional area $A$, and a 
piston height $z$, so that 
$$
V=Az,\ \  F=Af,\ \ S=As, \eqno(1.5)
$$
then the free energy per unit area obeys 
$$
df=-sdT-Pdz, \eqno(1.6)
$$
The second law of thermodynamics dictates that the isothermal 
compressibility  
$$
K_T=-\Big({1\over V}\Big)\Big({\partial V\over \partial P}\Big)_T =
-\Big({1\over z}\Big)\Big({\partial z\over \partial P}\Big)_T ,
\eqno(1.6)
$$
obeys the inequality 
$$
K_T\ge 0\ \ \ \ {\rm (second\ law)}. 
\eqno(1.7)
$$
It is this above positive compressibility implication of the 
second law of thermodynamics that is violated by the vacuum in 
the conventional quantum electrodynamic Casimir effect.

\bigskip 
\centerline{\bf 2. Statistical Thermodynamics of the Casimir Effect}
\medskip

Consider two thick parallel conducting plates separated by a 
distance $z$ with the ``vacuum'' material between the plates. 
Let $f(z,T)$ be the free energy per unit area of this vacuum. 
The ground state energy per unit area of this vacuum is given by 
[3-5]
$$
\epsilon (z)=\lim_{T\to 0}f(z,T)=-\Big({\pi^2\over 720}\Big)
\Big({\hbar c\over z^3}\Big), \eqno(2.1)
$$
yielding the pressure  
$$
P_0=\lim_{T\to 0}P(z,T)=-\Big({\pi^2\over 240}\Big)
\Big({\hbar c\over z^4}\Big). \eqno(2.2)
$$
The zero temperature compressibility then reads 
$$
K_0=\lim_{T\to 0}K_T=-\Big({60\over \pi^2 }\Big)
\Big({z^4 \over \hbar c }\Big)<0. \eqno(2.3)
$$
The negative compressibility ($K_0<0$) in Eq.(2.3) violates the 
second law of thermodynamics, as written in Eq.(1.7). 

\medskip 
\centerline{\bf 3. Stability and High Frequency Driving Forces}
\medskip

If in addition to the Casimir effect one applies a voltage source 
$U$ across the plates, then the total free energy per unit area obeys 
$$
df_{tot}=-sdT-Pdz-\sigma dU, \eqno(3.1)
$$  
where $\sigma $ represents that charge per unit area on the capacitor 
plates. The total free energy per unit area (in Gaussian units) for 
parallel plates with capacitance $C(z)=\{A/(4\pi z)\}$ obeys 
$$
f_{tot}(T,z,U)=f(T,z)-\Big({U^2\over 8\pi z}\Big). \eqno(3.2)
$$
The Coulomb attraction last term on the right hand side of Eq.(3.3) only 
serves to enhance the thermodynamic second law violation in the 
compressibility of the space between the plates. In fact, even if the 
Coulomb force dominates the Casimir force one appears to still have 
a second law violation merely from Coulomb law. 

Employing an experimental viewpoint, the situation becomes stable if the 
voltage has a high frequency AC as well as a DC component [6], 
$$
U=u+\sqrt{2}u_\omega \cos(\omega t). \eqno(3.3)
$$
After averaging over the high frequency force components, 
the effective energy per unit area ($T \to 0$) becomes  
$$
\bar{\epsilon }(z)=-\Big({\pi^2\over 720}\Big)
\Big({\hbar c\over z^3}\Big)
-\Big({u^2+u_\omega^2\over 8\pi z}\Big)
+\Big({K(\omega ,u,u_\omega )\over z^4}\Big), \eqno(3.4)
$$
where 
$$
K(\omega ,u,u_\omega )=\Big({1\over 512\pi ^2 }\Big)
\Big({u_\omega ^4+8u^2u_\omega ^2\over \mu \omega^2}\Big). 
\eqno(3.5)
$$
and $\mu $ represents the mass per unit area of the vibrating 
capacitor plate. 

Thermodynamic ``stability'' can be achieved at the equilibrium 
distance $z=Z$ which minimizes the energy $\bar{\epsilon }(z)$. 
This equilibrium does correspond to a positive compressibility 
$\bar{\epsilon }^{\prime \prime }(Z)>0$. However, it requires 
continual total entropy production to maintain such vibrating plate 
``equilibrium'' stability. This is not the normal sort of thing 
(say minimum energy at fixed entropy) thought 
to represent the equilibrium thermodynamic second law. 
However, such a high frequency technique might be useful 
experimentally.

\bigskip
\centerline{\bf 4. Conclusions}
\medskip

The notion of negative compressibility matter is as old as the 
van ter Waals approximation to the equations of state of a material [7]. 
 It has always been stated 
that such equations of state require supplimentary conditions such as 
equal area constructions and so forth (see for example Ref. 7 pp 257-62). 
Furthermore, the second 
thermodynamic law has been thought to put a complete and total 
veto on observing the totally unstable part of the van der Waals 
curve; i.e. $K_T<0$ exists formally in the approximation, but is 
strictly forbidden from observation.   

So now we have a paradox, and perhaps an interesting energy source. 
For the Casimir force, and even for Coulomb law, the regime 
$K_T<0$ may be asserted to be real. A closer study as to whether or 
not one can use such observable regimes to extract work in a cycle 
from an isothermal environment is underway.
\bigskip
 
\centerline{\bf References}
\medskip

\par \noindent
[1] \ \   Bekenstein J. D. 1973 {\it Phys. Rev. D} {\bf 7}, 2333.
\par 

\noindent [2] \ \   Hawking S. W. 1975 {\it Comm. Math. Phys.},
{\bf 43}, 199.\par

\noindent [3] \ \   Plunien G., B. Muller B., and  Greiner W.  
1986 {\it Phys. Rep.} {\bf 134}, 87.\par   

\noindent [4] \ \   Milonni P. W. 1994 
{\it The Quantum Vacuum} (London: Academic Press) pp 54-8.\par

\noindent [5] \ \   Mostepanenko V. M.  and  Trunov N. N. 1997 
{\it The Casimir Effect and its Applications} \par
\noindent \ \ \ \ \ \ (Oxford: Clarendon Press) pp 1-10. \par

\noindent [6] \ \  Landau L. D. and Lifshitz E. M. 1988 
{\it Mechanics} (Oxford: Pergamon Press) pp 93-5.  

\noindent [7] \ \  Landau L. D. and Lifshitz E. M. 1994 
{\it Statistical Physics I} (Oxford: Pergamon Press)\par
\noindent \ \ \ \ \ \ pp 232-5. 
\bye